\documentclass[a4paper,12pt]{article}
\usepackage{amsmath,enumerate}
\usepackage{graphicx}

\newcommand{\lapprox}{%
\mathrel{%
\setbox0=\hbox{$<$}
\raise0.6ex\copy0\kern-\wd0
\lower0.65ex\hbox{$\sim$}
}}
\newcommand{\gapprox}{%
\mathrel{%
\setbox0=\hbox{$>$}
\raise0.6ex\copy0\kern-\wd0
\lower0.65ex\hbox{$\sim$}
}}

\begin{document}

\begin{center}

{\Large \bf Neutrino masses and mixing angles in a model with six Higgs
triplets and $A_4$ symmetry}\\[20mm]

Raghavendra Srikanth Hundi\footnote{rshundi@phy.iith.ac.in} and
Itishree Sethi\footnote{ph15resch11004@iith.ac.in}\\
Department of Physics, Indian Institute of Technology Hyderabad,\\
Kandi - 502 285, India.\\[20mm]

\end{center}

\begin{abstract}

We have considered a model \cite{ma-weg}, where masses and a mixing pattern
for neutrinos
are governed by six Higgs triplets and $A_4$ symmetry. In this model
we have applied a certain diagonalisation procedure through which we have
shown that neutrino masses can have both normal or inverted hierarchy.
We have also shown that current neutrino oscillation data can be
explained in this model.

\end{abstract}

\newpage
\section{Introduction}

Neutrino masses and mixing angles play a vital role in our understanding
about physics beyond the standard model \cite{bsm}.
For a review on neutrino masses and mixing
angles, see ref.\cite{nu-rev}. One of the unknown facts about neutrino masses is
that we do not know how these masses have been ordered. Data from experiments
indicate that neutrino masses can be arranged in either normal or
inverted hierarchy \cite{nu-rev}. The problem related to neutrino mixing
angles is explained below.
From the fits to various neutrino oscillation data, three mixing angles and the
CP violating Dirac phase ($\delta_{\rm CP}$)
in the neutrino sector have been found \cite{glo-fit}. Out of the
three mixing angles, the values of $\theta_{12}$ and $\theta_{23}$ are
consistent
with $\sin^2\theta_{12}=1/3$ and $\sin^2\theta_{23}=1/2$, respectively. The
third mixing angle is small and it is found that
$\sin^2\theta_{13}\sim10^{-2}$ \cite{glo-fit}.
To a good approximation the three neutrino mixing angles are close
to the following pattern: $\sin^2\theta_{12}=1/3,\sin^2\theta_{23}=1/2,
\sin^2\theta_{13}=0$. This is known as Tribimaximal (TBM) mixing \cite{TBM}.
From
this we can infer that the mixing angles in the neutrino sector are not
arbitrary
but could emerge from a pattern. Based on this, one would like to know if
there is any underlying physics that is responsible for the pattern among the
neutrino mixing angles.

To address the above mentioned problem, several theoretical models based
on discrete symmetries have been proposed. For a review on these models and
related works, see refs.\cite{rev,relwo}. Out of these, models based on $A_4$
symmetry \cite{ma-raj,a4-mod} are elegant in explaining the mixing pattern in
the neutrino sector. Among these various models of $A_4$ symmetry, here
we particularly focus on one model \cite{ma-weg},
which is proposed by Ma and Wegman.
In this model, six Higgs triplets are
introduced along with the standard model (SM) fields \cite{ma-weg}.
Neutrinos, in this model, acquire non-zero masses through Type II seesaw
mechanism \cite{t2se}, where the neutral component of Higgs triplets get
vacuum expectation values (vevs).
By choosing certain $A_4$ symmetric charges for SM fields
and Higgs triplets, mixing
pattern among neutrinos has been explained in this model. Some details
related to these are given in the next section.

The above mentioned model is versatile, which was proposed soon after the
T2K Collaboration had found \cite{t2k}, for the first time,
that the mixing angle
$\theta_{13}$ is non-zero. This model has rich phenomenology, since
it has six Higgs triplets. One can study correlation between neutrino
oscillation observables and the phenomenology due to Higgs triplets in
this model. We discuss phenomenological implications of this model in
section 6. But before we study on that phenomenology, we have found
that there are few limitations about
the results obtained in ref.\cite{ma-weg}. In the work of ref.\cite{ma-weg},
results are obtained after assuming vevs
of some particular two Higgs triplets be equal and opposite. We elaborate
on this assumption in the next section where we briefly describe their work.
After making this assumption, one conclusion from the results of
ref.\cite{ma-weg} is that the neutrino masses in this
model can only be in normal hierarchy. In the present work, we have
analysed the same model as it is proposed in ref.\cite{ma-weg}, but we
make some assumptions about vevs of Higgs triplets
which are different from that in ref.\cite{ma-weg}. Following
from our assumptions, we have shown that not only normal but also inverted
hierarchy for neutrino masses is possible in this model.
Moreover, we have shown that this model is compatible with any currently
acceptable values for neutrino mixing angles and $\delta_{\rm CP}$.

In the model of ref.\cite{ma-weg}, after the six Higgs triplets get
vevs, neutrinos acquire a mixing mass matrix in the flavour basis.
This mass matrix should be diagonalised by a unitary matrix and from this
we can find the neutrino mixing angles and $\delta_{\rm CP}$.
In this work, in order to diagonalise
this mass matrix we develop an approximation scheme, after
making some assumptions about the vevs of the Higgs triplets.
From our approximation scheme, we obtain the leading order expressions for the
three neutrino mixing angles and $\delta_{\rm CP}$. The approximation scheme
that is applied in this work can have similarities with that in other works
of refs.\cite{ap-sc}.
But difference can be seen in the way the mixing angles and
$\delta_{\rm CP}$ are computed in our work as compared to that in other
works.

The paper is organised as follows. In the next section we describe the
model of ref.\cite{ma-weg}.
In section 3 we explain the assumptions we make in our work and describe
a procedure for diagonalising the mixing mass matrix for the neutrinos.
In section 4 we obtain leading order
expressions for the neutrino mixing angles and $\delta_{\rm CP}$. In section
5 we present numerical results of our work. In section 6 we describe the
phenomenological implications of the model of ref.\cite{ma-weg}.
We conclude in the last section.

\section{The model}

The model we consider is an extension of SM where the additional
fields are 2 extra Higgs doublets and 6 Higgs triplets \cite{ma-weg}.
In this model, $A_4$ symmetry is imposed in addition to the SM gauge symmetry.
The field content of this model in the neutrino sector and also their
charge assignments under $A_4$ and electroweak symmetries are given in table 1.
\begin{table}[h]
\centering
\begin{tabular}{|c|c|c|c|c|c|c|c|c|c|} \hline
Field & $L_i$ & $\ell_1^c$ & $\ell_2^c$ & $\ell_3^c$ & $\Phi_i$ & $\xi_1$
& $\xi_2$ & $\xi_3$ & $\xi_j$ \\ \hline
$A_4$ & $\underline{3}$ & $\underline{1}$ & $\underline{1}^\prime$ &
$\underline{1}^{\prime\prime}$ & $\underline{3}$ & $\underline{1}$ &
$\underline{1}^\prime$ & $\underline{1}^{\prime\prime}$ & $\underline{3}$
\\ \hline
$SU(2)_L$ & 2 & 1 & 1 & 1 & 2 & 3 & 3 & 3 & 3 \\ \hline
$U(1)_Y$ & $-\frac{1}{2}$ & $-1$ & $-1$ & $-1$ & $\frac{1}{2}$ & 1 & 1 & 1 & 1
\\ \hline
\end{tabular}
\caption{Relavant fields in the neutrino sector in the model of
ref.\cite{ma-weg}. Charge assignments of these fields under $A_4$ and
electroweak symmetries are also given. Here, $i=1,2,3$ and $j=4,5,6$.}
\end{table}
$A_4$ has the following 4 irreducible representations:
$\underline{1},\underline{1}^\prime,\underline{1}^{\prime\prime},
\underline{3}$. Under $A_4$, $SU(2)_L$ doublets and singlets of leptons are
assigned as: $L_i=(\nu_i,\ell_i)\sim\underline{3},\ell^c_1\sim\underline{1},
\ell^c_2\sim\underline{1}^\prime,\ell^c_3\sim\underline{1}^{\prime\prime}$.
Here, $i=1,2,3$. In the above mentioned model, altogether there are 3 Higgs
doublets which we denote them as $\Phi_i,i=1,2,3$. These doublets are
assigned under $\underline{3}$ of $A_4$. With these charge assignments, the
Yukawa couplings for charge leptons can be written as \cite{ma-raj}
\begin{equation}
{\cal L}=h_{ijk}\overline{L_i}\ell_j^c\Phi_k+h.c.
\end{equation}
Here, $i,j,k=1,2,3$. $h_{ijk}$ are Yukawa coupings, whose form is determined
by $A_4$ symmetry, which can be seen in ref.\cite{ma-raj}. Assuming that
the 3 Higgs doublets acquire the same vev after the electroweak symmetry
breaking, we get a mixing mass matrix for charged leptons. This mass matrix
can be diagonalized with the following transformations on the charged lepton
fields \cite{ma-raj}.
\begin{eqnarray}
&& \Psi_L\to U_L\Psi_L,\quad \Psi_R\to U_R\Psi_R,
\nonumber \\
&& \Psi_L=(\ell_1,\ell_2,\ell_3)^{\rm T},\quad
\Psi_R=(\ell^c_1,\ell^c_2,\ell^c_3)^{\rm T},
\nonumber \\
&& U_L = U_{CW} = \frac{1}{\sqrt{3}}\left(\begin{array}{ccc}
1 & 1 & 1 \\
1 & \omega & \omega^2 \\
1 & \omega^2 & \omega \end{array}\right),\quad
U_R = \left(\begin{array}{ccc}
1 & 0 & 0\\
0 & 1 & 0\\
0 & 0 & 1 \end{array}\right).
\end{eqnarray}
Here, $\omega=e^{2\pi i/3}$.

As stated before that neutrinos in this model acquire masses through Type II
seesaw mechanism \cite{t2se}, when the 6 Higgs triplets get
vevs. Denoting these 6 Higgs triplets as $\xi_i,i=1,\cdots,6$, under $A_4$
their charges are assigned as follows: $\xi_1\sim\underline{1},\xi_2\sim
\underline{1}^\prime,\xi_3\sim\underline{1}^{\prime\prime},
\xi_j\sim\underline{3}$.
Here, $j=4,5,6$. After these Higgs triplets get vevs, mass terms for neutrinos
can be written as follows \cite{ma-weg}.
\begin{eqnarray}
{\cal L} &=& \overline{\Psi^c}_\nu{\cal M}_\nu\Psi_\nu + h.c.,
\quad
\Psi_\nu = (\nu_1,\nu_2,\nu_3)^{\rm T},\quad
\Psi^c_\nu=C\bar{\Psi}_\nu^{\rm T},
\nonumber \\
{\cal M}_\nu &=& \left(\begin{array}{ccc}
a+b+c & f & e\\
f & a+\omega b+\omega^2c & d\\
e & d & a+\omega^2b+\omega c \end{array}\right).
\label{eq:Mnu}
\end{eqnarray}
Here, $C$ is the charge conjugation matrix. In the above equation,
$a,b,c,d,e,f$ come from $\langle\xi_1^0\rangle,\langle\xi_2^0\rangle,
\langle\xi_3^0\rangle,\langle\xi_4^0\rangle,\langle\xi_5^0\rangle,
\langle\xi_6^0\rangle$, respectively \cite{ma-weg}. After applying the
following transformation on $\Psi_\nu$ as
\begin{equation}
\Psi_\nu\to U_{CW}U_{\rm TBM}\Psi_\nu,\quad
U_{\rm TBM} = \left(\begin{array}{ccc}
\sqrt{2/3} & 1/\sqrt{3} & 0 \\
-1/\sqrt{6} & 1/\sqrt{3} & -1/\sqrt{2} \\
-1/\sqrt{6} & 1/\sqrt{3} & 1/\sqrt{2} \end{array}\right),
\label{eq:trans}
\end{equation}
the matrix ${\cal M}_\nu$ of Eq. (\ref{eq:Mnu}) would transform to
\begin{equation}
{\cal M}^\prime_\nu = \left(\begin{array}{ccc}
a-(b+c)/2+d & (f+e)/\sqrt{2} & (b-c)\sqrt{3}/2 \\
(f+e)/\sqrt{2} & a+b+c & i(e-f)/\sqrt{2} \\
(b-c)\sqrt{3}/2 & i(e-f)/\sqrt{2} & -a+(b+c)/2+d \end{array}\right)
\label{eq:Mnu2}
\end{equation}
The above matrix would be in diagonal form if $e=f=0$ and $b=c$ and in this
case, from the transformations of charged leptons and neutrinos, we can
notice that $U_{\rm TBM}$ is the unitary matrix which diagonalises the
neutrino mass matrix in a basis where charged lepton masses are
already diagonalised. Hence $U_{\rm TBM}$ can be identified as the
Pontecorvo-Maki-Nakagawa-Sakata (PMNS) matrix. We can parametrise the
PMNS matrix ($U_{\rm PMNS}$) in terms of neutrino mixing
angles, which we have given in section 4. After equating $U_{\rm TBM}$
with $U_{\rm PMNS}$ we can find that the neutrino mixing angles fit the
TBM pattern, which we have described in the previous section.
But in the above mentioned case, where $e=f=0$ and $b=c$ , the angle
$\theta_{13}$ would
become zero and this possibility is ruled out by the oscillation data. Hence,
in order to get $\theta_{13}\neq 0$, at least some of $e$, $f$ and
$b-c$ should have non-zero values.

Based on the observations made in the previous paragraph, in ref.\cite{ma-weg},
$\theta_{13}$ has been shown to be non-zero by assuming $e=-f\neq 0$ and
$b-c\neq0$. But by considering this possibility it has been concluded that
the neutrinos can only have normal mass hierarchy. Although we
should assume $e$ and $f$ to be non-zero, in general there need not be
any constraint between them. In this work we consider non-zero values
for $e$, $f$ and $b-c$, but otherwise do not assume any relation between
$e$ and $f$.

\section{Diagonalisation procedure and neutrino masses}

In this section we explain our methodology of diagonalising the matrix
${\cal M}_\nu$ of
Eq. (\ref{eq:Mnu}). As explained in the previous section that after
applying the transformation of Eq. (\ref{eq:trans}) on
${\cal M}_\nu$ of Eq. (\ref{eq:Mnu}),
we have got the mixing mass matrix among neutrinos which is given by
${\cal M}^\prime_\nu$. We can notice that ${\cal M}^\prime_\nu$
is nearly diagonal if we assume $e$, $f$ and $b-c$ are small values. After
assuming that these are small, we can expect that ${\cal M}^\prime_\nu$
can be diagoanlised by a unitary matrix which is nearly
equal to unit matrix. This unitary matrix can be parametrised, upto
first order, as \cite{ap-sc}
\begin{equation}
U_\epsilon = \left(\begin{array}{ccc}
1 & \epsilon_{12} & \epsilon_{13} \\
-\epsilon_{12}^* & 1 & \epsilon_{23} \\
-\epsilon_{13}^* & -\epsilon_{23}^* & 1 \end{array}\right)
\label{eq:uep}
\end{equation}
In the above equation, $\epsilon_{12},\epsilon_{13},\epsilon_{23}$ are small
and complex.

In the above described methodology, in order to diagonalise the
matrix ${\cal M}_\nu$ of Eq. (\ref{eq:Mnu}), we are applying
the following transfromation on the neutrino fields
\begin{equation}
\Psi_\nu\to U_{CW}U_{\rm TBM}U_\epsilon\Psi_\nu
\label{eq:trans2}
\end{equation}
Now, from the transformations of charged leptons and neutrinos,
we can notice that the PMNS matrix in this model would be
\begin{equation}
U_{\rm PMNS} = U_{\rm TBM}U_\epsilon
\label{eq:pmns}
\end{equation}
As explained before that $U_{\rm PMNS}$ can be parametrised in terms of
neutrino mixing angles.
Hence from the above relation we may hope to get $\theta_{13}$ to be non-zero
for some particular values of $\epsilon$-parameters. As mentioned
before that these $\epsilon$-parameters need to be small, since in our
diagonalisation procedure we have assumed that $e$, $f$ and $b-c$ of
${\cal M}^\prime_\nu$ should be small.
Here we quantify how small these parameters need to be. As mentioned
previously that the neutrino oscillation data predicts that
$\sin^2\theta_{13}\approx2\times 10^{-2}$ which is very small in comparision
to unity. So we can take $\sin\theta_{13}\approx0.15$ to be a small value.
Based on this observation,
we assume that the real and imaginary parts of $\epsilon$-parameters
to be atmost of the order of $\sin\theta_{13}$. By making this
assumption we show later that we get consistent results in our work.

As explained previously that we are applying the transformation of Eq.
(\ref{eq:trans2}) on ${\cal M}_\nu$ of Eq. (\ref{eq:Mnu}).
As a result of this, we can notice that, effectively the matrix
${\cal M}^\prime_\nu$ is diagonalised by $U_\epsilon$. Relation for the
diagonalisation of ${\cal M}^\prime_\nu$ can be expressed as
\begin{equation}
{\cal M}^\prime_\nu = U_\epsilon^*\cdot{\rm diag}(m_1,m_2,m_3)\cdot
U_\epsilon^\dagger
\label{eq:mdia}
\end{equation}
Here, $m_1,m_2,m_3$ are masses of neutrinos. Neutrino masses
can be estimated from the global fits to the neutrino oscillation data
\cite{glo-fit}. From these global fits we know that there are two
mass-square differences among the neutrino masses, which are given below
\cite{glo-fit}.
\begin{eqnarray}
&& m_{\rm sol}^2 = m_2^2-m_1^2 = 7.39\times 10^{-5}~{\rm ev}^2,
\nonumber \\
&& m_{\rm atm}^2 = \left\{\begin{array}{l}
m_3^2-m_1^2 = +2.525\times 10^{-3}~{\rm eV}^2\quad({\rm normal~hierarchy}) \\
m_3^2-m_2^2 = -2.512\times 10^{-3}~{\rm eV}^2\quad({\rm inverted~hierarchy})
\end{array}\right.
\end{eqnarray}
In the above we have given the best fit values. Here $m_{\rm sol}$ and
$m_{\rm atm}$ represent solar and atmospheric mass scales respectively.
To fit the
above mass-square differences we can take neutrino masses as
\begin{eqnarray}
&& m_1\lapprox m_{\rm sol},\quad m_2=\sqrt{m_1^2+m_{\rm sol}^2},\quad
m_3=\sqrt{m_1^2+m_{\rm atm}^2}\quad({\rm NH})
\nonumber \\
&& m_3\lapprox m_{\rm sol},\quad m_2=\sqrt{m_3^2-m_{\rm atm}^2},\quad
m_1=\sqrt{m_2^2-m_{\rm sol}^2}\quad({\rm IH})
\end{eqnarray}
Here, NH(IH) indicate normal(inverted) hierarchy. In the case of IH, by
taking $m_3=m_{\rm sol}$ we would get $\sum m_\nu=m_1+m_2+m_3=0.11$ eV. This
value is just below the upper bound on the sum of neutrino masses obtained
by Planck, which is 0.12 eV \cite{planck}.
On the other hand, in the case of NH, even
if we take $m_1=m_{\rm sol}$ we would get $\sum m_\nu=0.07$ eV, which is
reasonably below the above mentioned upper bound.

In the diagonalisation procedure that we have described above,
to find the neutrino masses
we need to solve the relations in Eq. (\ref{eq:mdia}). We can notice
here that the matrix ${\cal M}^\prime_\nu$ contain all the model parameters
related to neutrino masses. From Eq. (\ref{eq:mdia}) it is clear that these
model parameters are related to mass eigenvalues of neutrinos and
$\epsilon$-parameters. In the next section we will show that these
$\epsilon$-parameters can be determined from the neutrino mixing angles and
$\delta_{\rm CP}$, whose values are found the oscillation data \cite{glo-fit}.
As for the mass eigenvalues of neutrinos we have described above that they
be chosen from mass-square differences which are also found from
the oscillation data. Now,
after using Eq. (\ref{eq:mdia}) we can proceed to
find the model parameters of ${\cal M}^\prime_\nu$ in terms of
observables from oscillation data. Before doing that let us mention that
the oscillation data predict
that there is a hierarchy between the two neutrino mass-square differences.
In fact, from the global fits to oscillation data, we can notice that
$\frac{m_{\rm sol}^2}{m_{\rm atm}^2}\sim\sin^2\theta_{13}\sim10^{-2}$
\cite{glo-fit}.
As mentioned previously, quantities which are of the
order of $\frac{m_{\rm sol}^2}{m_{\rm atm}^2}$ or $\sin^2\theta_{13}$ are
very small in comparision to unity and so we neglect them
in our analysis. As a result of this, we compute terms which are of upto
first order in
$\sin\theta_{13}\sim\frac{m_{\rm sol}}{m_{\rm atm}}$, in the
right hand side of Eq. (\ref{eq:mdia}).
We do this computation in both the cases of NH and IH. In either of
these cases, the mass eigenvalues of neutrinos in terms of model parameters
are found to be same, which are given below.
\begin{equation}
m_1=a+d-\frac{b+c}{2},\quad m_2=a+b+c,\quad m_3=-a+d+\frac{b+c}{2}.
\label{eq:rel1}
\end{equation}
Whereas, relations for other model paramteres are found to be dependent
on neutrino mass hierarchy. These relations are given below.
\begin{eqnarray}
&& {\rm NH}:\quad e+f=0,\quad \frac{\sqrt{3}}{2}(b-c)=m_3\epsilon^*_{13},
\quad \frac{i}{\sqrt{2}}(e-f)=m_3\epsilon^*_{23}.
\nonumber \\
&& {\rm IH}:\quad \frac{e+f}{\sqrt{2}}=-m_1\epsilon_{12}+m_2\epsilon^*_{12},
\quad \frac{\sqrt{3}}{2}(b-c)=-m_1\epsilon_{13},\quad
\frac{i}{\sqrt{2}}(e-f)=-m_2\epsilon_{23}.
\nonumber \\
\label{eq:rel2}
\end{eqnarray}
Using the above relations we can see that the diagonal elements of the matrix
${\cal M}^\prime_\nu$, up to first order approximation, would be same as
the mass
eigenvalues of neutrinos. Whereas, the off-diagonal elements in
${\cal M}^\prime_\nu$ are related to neutrino masses and $\epsilon$-parameters.
Previously we have assumed that the real and imaginary parts of
$\epsilon$-parameters to be around $\sin\theta_{13}$. As a result of this,
the relations in Eq. (\ref{eq:rel2}) suggest that the off-diagonal elements
of the matrix ${\cal M}^\prime_\nu$ are suppressed by
${\cal O}(\sin\theta_{13})$ as compared to the neutrino mass eigenvalues.
This result is consistent with the asumption we made before that
$e$, $f$ and $b-c$ should be small values.

Using the realtions of Eqs. (\ref{eq:rel1}) $\&$ (\ref{eq:rel2}),
depending on the case of NH or IH, we can determine all the
model parameters in terms
of neutrino mass eigenvalues and the $\epsilon$-parameters. As stated
previously that these $\epsilon$-parameters can be found from
the neutrino mixing angles and $\delta_{\rm CP}$, which is the subject
of the next section. So we can state that
by appropriately choosing the model parameters we can explain either the
normal or inverted hierarchy mass spectrum for neutrinos in
this model.
Here it is worth to mention that in the case of NH,
we have $e=-f$. This is exactly what it is assumed in ref.\cite{ma-weg}
and as a result of this it has been concluded that neutrinos can only have
normal mass hierarchy. So
our results are agreeing with that of ref.\cite{ma-weg} in the case of NH.
But in addition
to this, we have shown that inverted mass hierarchy for neutrinos can
also be possible in this model.

\section{Neutrino mixing angles}

In the previous section we have explained that in order to get $\theta_{13}$
to be non-zero, we have chosen to follow a certain diagonalisation
procedure through
which we have shown that the PMNS matrix in our model could be given
by Eq. (\ref{eq:pmns}). The PMNS matrix can be parametrised in terms
of neutrino mixing angels and a Dirac CP-violating phase, $\delta_{\rm CP}$.
After using this parametrisation in Eq. (\ref{eq:pmns}) we can get
relations among neutrino mixing angles, $\delta_{\rm CP}$ and
$\epsilon$-parameters. In this section, we will solve these relations and
show that all the three neutrino mixing angles get deviations away from
the TBM pattern and hence $\theta_{13}\neq 0$.

To express the PMNS matrix in terms of neutrino mixing angles and
$\delta_{\rm CP}$, we follow the PDG convention, which we have
given below \cite{pdg}.
\begin{equation}
U_{\rm PMNS} = \left(\begin{array}{ccc}
c_{12}c_{13} & s_{12}c_{13} & s_{13}e^{-i\delta_{\rm CP}} \\
-s_{12}c_{23}-c_{12}s_{23}s_{13}e^{i\delta_{\rm CP}} &
c_{12}c_{23}-s_{12}s_{23}s_{13}e^{i\delta_{\rm CP}} & s_{23}c_{13} \\
s_{12}s_{23}-c_{12}c_{23}s_{13}e^{i\delta_{\rm CP}} &
-c_{12}s_{23}-s_{12}c_{23}s_{13}e^{i\delta_{\rm CP}} & c_{23}c_{13}
\end{array}\right)
\label{eq:pmns2}
\end{equation}
Here, $c_{ij}=\cos\theta_{ij}$ and $s_{ij}=\sin\theta_{ij}$.
We use the above form of $U_{\rm PMNS}$ in Eq. (\ref{eq:pmns}) and
determine the neutrino mixing angles and $\delta_{\rm CP}$ in terms of
$\epsilon$-parameters. Since
these $\epsilon$-parameters are complex we can write them as
\begin{equation}
\epsilon_{ij} = Re(\epsilon_{ij})+i Im(\epsilon_{ij}),\quad i,j=1,2,3.
\end{equation}
Here, $Re(\epsilon_{ij})$ and $Im(\epsilon_{ij})$ are real and imaginary
parts of $\epsilon_{ij}$.

As explained above that we use the form for $U_{\rm PMNS}$ of Eq.
(\ref{eq:pmns2}) in Eq. (\ref{eq:pmns}). After equating the 13-elements
in the matrix
relation of Eq. (\ref{eq:pmns}), we get the following relation for
$\sin\theta_{13}$.
\begin{equation}
s_{13}=\left(\sqrt{\frac{2}{3}}\epsilon_{13}+\frac{1}{\sqrt{3}}\epsilon_{23}
\right)e^{i\delta_{\rm CP}}.
\end{equation}
Since the sine of an angle is real, we need to demand that the imaginary part of
the right hand side of the above relation should be zero.
After doing this we get
\begin{eqnarray}
&& s_{13}=\left(\sqrt{\frac{2}{3}}Re(\epsilon_{13})+\frac{1}{\sqrt{3}}
Re(\epsilon_{23})\right)\cos\delta_{\rm CP} -
\left(\sqrt{\frac{2}{3}}Im(\epsilon_{13})+\frac{1}{\sqrt{3}}
Im(\epsilon_{23})\right)\sin\delta_{\rm CP}.
\nonumber \\ \label{eq:s13} \\
&& \left(\sqrt{\frac{2}{3}}Re(\epsilon_{13})+\frac{1}{\sqrt{3}}
Re(\epsilon_{23})\right)\sin\delta_{\rm CP} +
\left(\sqrt{\frac{2}{3}}Im(\epsilon_{13})+\frac{1}{\sqrt{3}}
Im(\epsilon_{23})\right)\cos\delta_{\rm CP} = 0.
\nonumber \\
\end{eqnarray}
From the above two equations we can see that both $\sin\theta_{13}$ and
$\delta_{\rm CP}$ can be determined in terms of $\epsilon_{13}$ and
$\epsilon_{23}$ parameters. Hence, by choosing some particular values for these
$\epsilon$-parameters we may hope to get consistent values for $\sin\theta_{13}$
and $\delta_{\rm CP}$. We present these numerical results on
$\epsilon$-parameters in the
next section. But before doing that we will apply the above described method
to obtain expressions for other sine of the angles, which is explained below.

As stated before that we are
neglecting terms of the order of $s_{13}^2$ in comparision to unity,
hence we have $c_{13}=\sqrt{1-s_{13}^2}=1+{\cal O}(s_{13}^2)\approx 1$.
Now that we have known $c_{13}$, by equating
12- and 23-elements of the matrix relation of Eq. (\ref{eq:pmns}), we can
determined $s_{12}$ and $s_{23}$ in terms of $\epsilon$-parameters. Here
again we need to demand that the sine of an angle should be real. After
doing this we get the following relations.
\begin{eqnarray}
&& s_{12}=\frac{1}{\sqrt{3}}+\sqrt{\frac{2}{3}}Re(\epsilon_{12}),\quad
s_{23}=-\frac{1}{\sqrt{2}}-\frac{1}{\sqrt{6}}Re(\epsilon_{13})+
\frac{1}{\sqrt{3}}Im(\epsilon_{23}),
\label{eq:s12,23}
\\
&& Im(\epsilon_{12})=0,\quad Im(\epsilon_{13})=\sqrt{2}
Im(\epsilon_{23})
\label{eq:cond}
\end{eqnarray}

In the above we have shown that the sine of the three neutrino mixing angles and
$\delta_{\rm CP}$ can be obtained in terms of $\epsilon$-parameters after
equating the 12-, 13- and 23-elements of the matrix relation of
Eq.(\ref{eq:pmns}). In our analysis we have three complex
$\epsilon$-parameters, whose real and imaginary parts will give us six
independent parameters. But from Eq. (\ref{eq:cond}) we can see that
$Im(\epsilon_{13})$ and $Im(\epsilon_{23})$ are not independent parameters and
$Im(\epsilon_{12})=0$. As a result of this the following four parameters can
be used to determine the three neutrino mixing angles and $\delta_{\rm CP}$:
$Re(\epsilon_{12})$,
$Re(\epsilon_{13})$, $Re(\epsilon_{23})$ and $Im(\epsilon_{13})$.

In the matrix relation of Eq.(\ref{eq:pmns}) we have equated 12-, 13- and
23-elements and found relations for the three neutrino mixing angles and
$\delta_{\rm CP}$ in terms of $\epsilon$-parameters. By now we have used
all the available $\epsilon$-parameters in determining the neutrino mixing
angles and $\delta_{\rm CP}$. These relations for neutrino mixing
angles and $\delta_{\rm CP}$ can be used in other elements of the matrix
relation of Eq. (\ref{eq:pmns}) and then we may expect to get
some constraints
among the $\epsilon$-parameters. Below we will demonstrate that no constraints
among these $\epsilon$-parameters will happen. Let us equate the
11-elements of the matrix relation of Eq. (\ref{eq:pmns}) and this would lead to
\begin{equation}
c_{12}c_{13}=\sqrt{\frac{2}{3}}-\frac{1}{\sqrt{3}}\epsilon^*_{12}.
\end{equation}
We can check that the above relation is satisfied self consistently upto
first order in $\epsilon$-parameters, after using Eqs. (\ref{eq:s13}),
(\ref{eq:s12,23}) $\&$ (\ref{eq:cond}). Similarly we have checked that the
relations we would get by equating
other elements of the matrix relation of Eq. (\ref{eq:pmns}) are satisfied
self consistenly upto first order in $\epsilon$-parameters after using
Eqs. (\ref{eq:s13}) - (\ref{eq:cond}). As a result of this, we do not get
any additional constraints on the $\epsilon$-parameters.

\section{Results}

In the previous section we have explained that the three neutrino mixing
angles and $\delta_{\rm CP}$ can be determined by $Re(\epsilon_{12})$,
$Re(\epsilon_{13})$, $Re(\epsilon_{23})$ and $Im(\epsilon_{13})$. In this
section we will show that for some particular values of these
$\epsilon$-parameters, the three neutrino mixing angles and
$\delta_{\rm CP}$ can be fitted to the observed values as found from
the oscillation
data \cite{glo-fit}. For this purpose in table 2 we mention the $3\sigma$
ranges obtained in the cases of NH and IH for the neutrino
mixing angles and $\delta_{\rm CP}$.
\begin{table}[h]
\centering
\begin{tabular}{|l|c|c|} \hline
 & NH & IH \\ \hline
$\sin^2\theta_{12}$ & 0.275$\to$0.350 & 0.275$\to$0.350 \\
$\sin^2\theta_{23}$ & 0.418$\to$0.627 & 0.423$\to$0.629 \\
$\sin^2\theta_{13}$ & 0.02045$\to$0.02439 & 0.02068$\to$0.02463 \\
$\delta_{\rm CP}/^o$ & 125$\to$392 & 196$\to$360 \\ \hline
\end{tabular}
\caption{$3\sigma$ ranges in the cases of both NH and IH for the square of
the sine of the three neutrino
mixing angles and the CP-violating Dirac phase \cite{glo-fit}.}
\end{table}

From the relations of Eq. (\ref{eq:s13}) - Eq. (\ref{eq:cond}), we can
obtain all $\epsilon$-parameters in terms of
neutrino mixing angles and $\delta_{\rm CP}$. Using the 3$\sigma$ range
for $\sin^2\theta_{12}$, we found the allowed range for $Re(\epsilon_{12})$
as: -6.19$\times10^{-2}$ to 1.77$\times10^{-2}$. We can see that the
magnitude of these allowed values are below $s_{13}\approx0.15$.
From the 3$\sigma$ ranges of $\sin^2\theta_{13}$, $\sin^2\theta_{23}$
and $\delta_{\rm CP}$ we can get allowed regions for
$Re(\epsilon_{13})$, $Re(\epsilon_{23})$ and $Im(\epsilon_{13})$.
These allowed regions are plotted in figure 1 in the case of NH.
\begin{figure}[!h]
\begin{center}

\includegraphics[height=3.0in,width=3.0in]{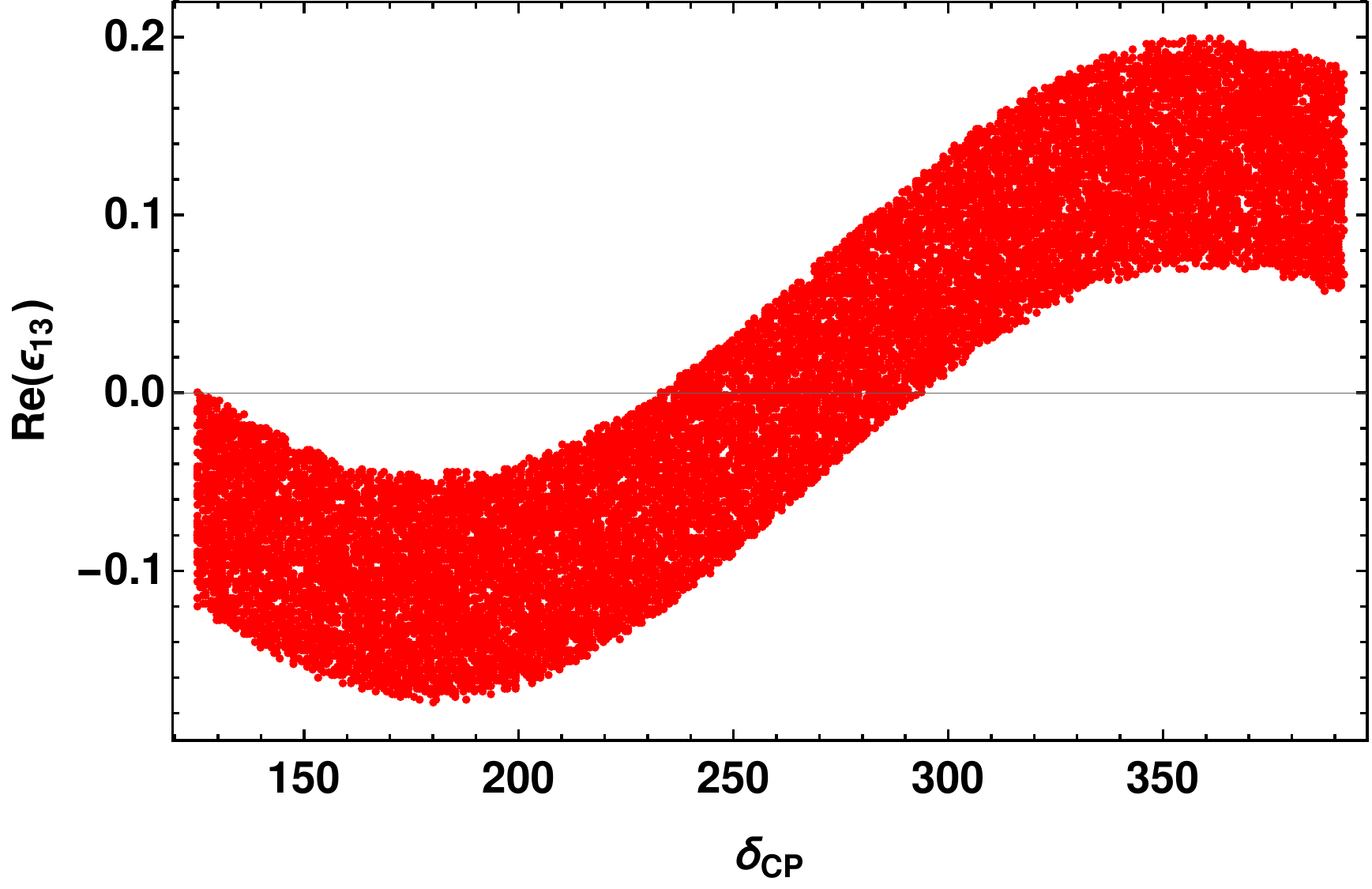}
\includegraphics[height=3.0in,width=3.0in]{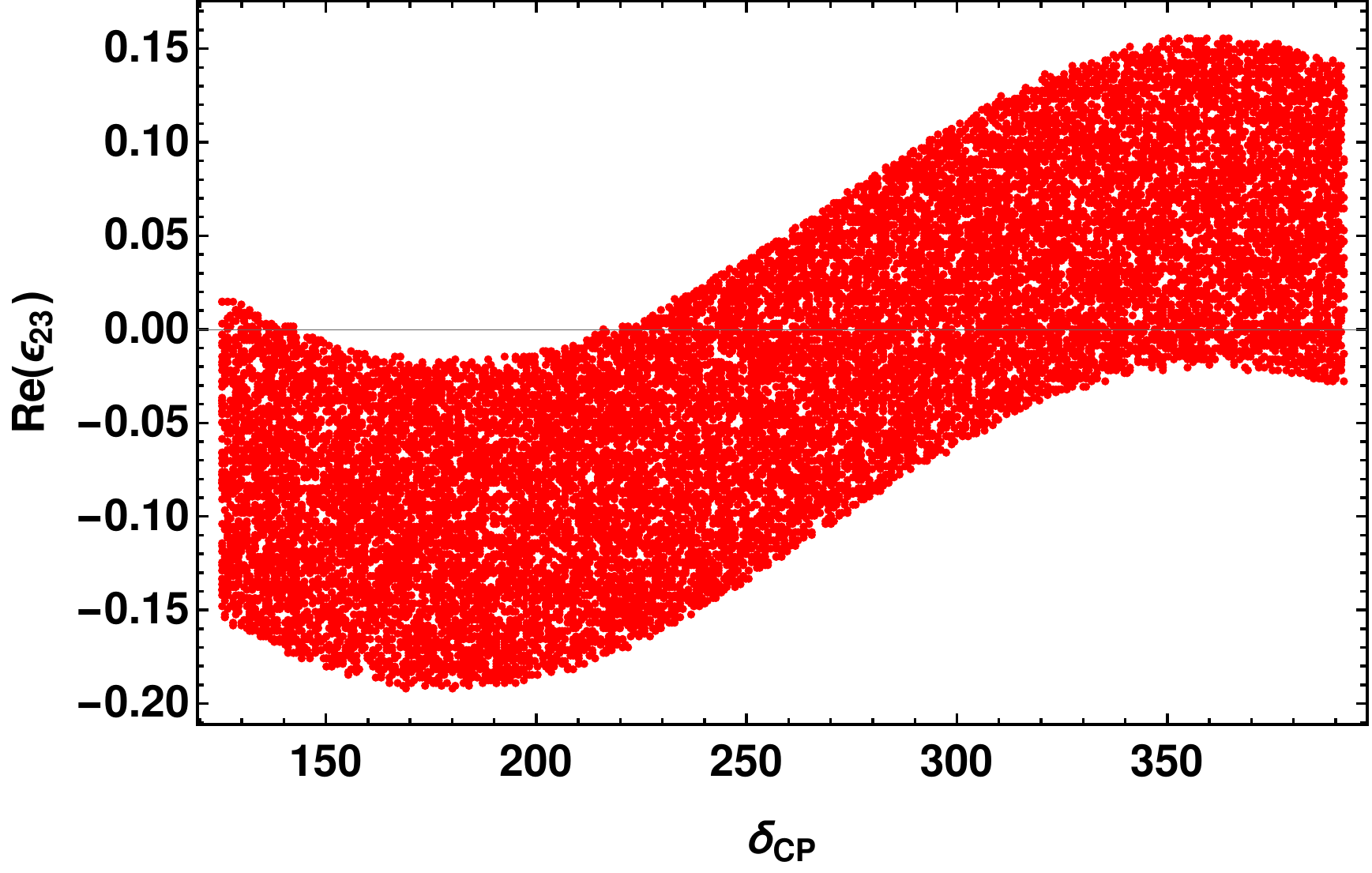}

\includegraphics[height=3.0in,width=3.0in]{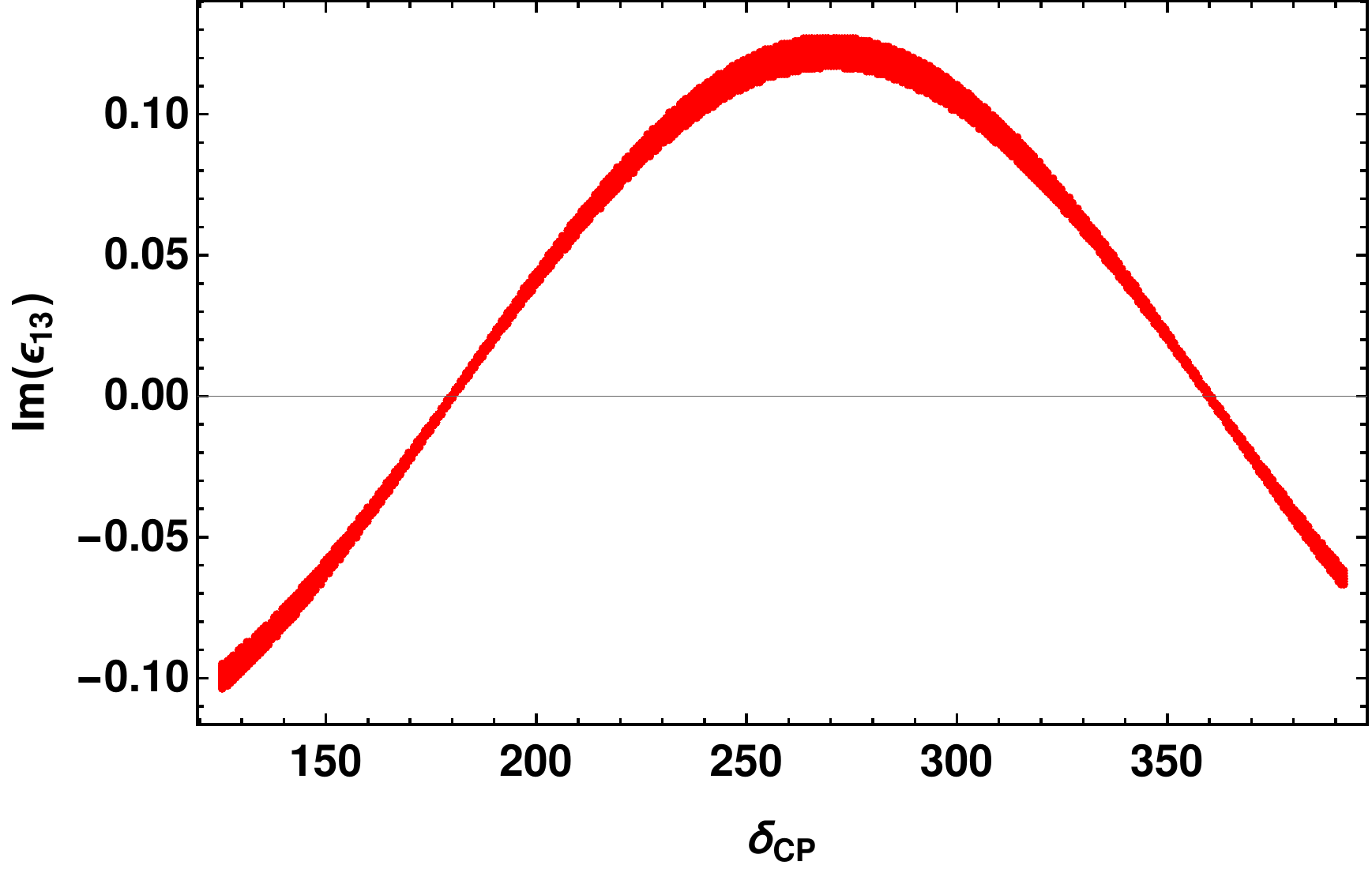}

\end{center}
\caption{Allowed regions for $Re(\epsilon_{13})$, $Re(\epsilon_{23})$ and
$Im(\epsilon_{13})$ are shown in the case of NH. $\delta_{\rm CP}$ is
expressed in degrees. In all the above
plots, 3$\sigma$ ranges for $\sin^2\theta_{13}$ and $\sin^2\theta_{23}$
have been used.}
\end{figure}
From this figure we can see that the values for $|Re(\epsilon_{13})|$
and $|Re(\epsilon_{23})|$ can be atmost of 0.2, which is just at the order of
$s_{13}\approx0.15$. In fact, $|Re(\epsilon_{13})|$
and $|Re(\epsilon_{23})|$ get maximum values when $\delta_{\rm CP}$
is around 180$^o$
or 360$^o$. Otherwise, these parameters can take values even less than 0.2.
As for the $|Im(\epsilon_{13})|$, we can
notice from figure 1 that this parameter can take a maximum of 0.13 when
$\delta_{\rm CP}$ is around 270$^o$.

We can notice from table 2 that the 3$\sigma$ ranges for the neutrino
mixing angles do not change much between NH and IH cases. The only significant
difference is that $\delta_{\rm CP}$ has a narrow allowed region in the
case of IH as compared that of NH. Because of this, we can expect that
the numerical limits quoted for $Re(\epsilon_{12})$, $Re(\epsilon_{13})$,
$Re(\epsilon_{23})$ and $Im(\epsilon_{13})$ in the case of NH would
almost be the same even in the case of IH. This we have seen after
computing the above mentioned parameters in the case of IH. In fact,
we have found that the allowed regions shown in figure 1 do not change
significantly in the case of IH, except for the fact that in IH the axis of
$\delta_{\rm CP}$ varies from 196$^o$ to 360$^o$.

From the numerical results described above we can see that all the
$\epsilon$-paramters, in the case of NH and IH, are less than or of the order
of $s_{13}$. This justifies the assumption we have made for diagoanlising
the neutrino mass matrix in section 3. This justification also vindicate
one of our results that both NH and IH cases are possible in the model of
ref.\cite{ma-weg}. Here we comment on the fact that the calculations done
in this work are upto first order in $s_{13}$. By including second order
terms we expect the relations mentioned in Eqs. (\ref{eq:rel1}) -
(\ref{eq:rel2}) $\&$ (\ref{eq:s13}) - (\ref{eq:cond}) get corrections
with terms which are of ${\cal O}(s_{13}^2)$. Since these second order
terms contribute very small values in the numerical analysis, we do not
expect any changes in the qualitative conclusions made in this work.

\section{Phenomenological implications of the model}

As stated in section 1 that neutrinos in the model of ref.\cite{ma-weg}
acquire masses through Type II seesaw mechanism. Hence, in this model,
lepton number is violated by two units and the neutrinos are Majorana
particles. As a result of this, one implication of this model is the
existance of neutrinoless double-beta decay. The rate of this decay
is related to effective Majorana mass, which is given below
\begin{equation}
m_{ee}=\left|\sum_{i=1}^3m_iU_{ei}^2\right|
\end{equation}
Here, $U_{ei}$ are elements in the first row of PMNS matrix, which
is given in section 4. So far the above mentioned decay has never been
observed in experiments and as a result of that the following upper bound on
$m_{ee}$ has been set: 61 $-$ 165 meV \cite{kzen}. In the expression for
$m_{ee}$, $m_i$ indicate the three mass eigenvalues of neutrinos. In our
analysis, these mass eigenvalues are related to model parameters through
Eq. (\ref{eq:rel1}). The elements $U_{ei}$ depend on neutrino mixing angles
and $\delta_{\rm CP}$. Using our results obtained in section 4 we can express
$U_{ei}$ in terms of $\epsilon$-parameters, which are related to
model parameters via Eq. (\ref{eq:rel2}). Hence, in our work, the
quantity $m_{ee}$ can be expressed in terms of model parameters. Using the
above mentioned fact that $m_{ee}$ has an upper bound from experiments, we
can get constraints on model parameters in both NH and IH cases. We study
these constraints in our future work.

In a Type II seesaw mechanism \cite{t2se}, charge lepton flavour violating
decays such as $\mu\to 3e$ and $\mu\to e\gamma$ are driven by charged
components of scalar triplet Higgs \cite{lfv-d}. These decays happen due to
Yukawa couplings of triplet Higss with lepton doublets.
Since in the model of ref.\cite{ma-weg},
Type II seesaw mechanism is responsible for neutrino mass generation, one
can expect the above mentioned flavour violating decays to happen in this
model as well. We have seen that the structure with six triplet Higgses
of this model
can explain the consistent neutrino mixing pattern. Now,
these triplet Higgses can also drive the above mentioned flavour violating
decays.
Hence, in this model there can exist a correlation between neutrino
mixing angles and the flavour violating decays. These flavour violating
decays are not observed in experiments and hence the branching ratios of these
decays are bounded from above \cite{pdg}. Using these experimental
contraints one can study the bounds on the masses of triplet Higgses. We
can expect that these bounds may depend on the neutrino mixing angles, since
there is a correlation between neutrino oscillation observables and the
decay rates of these flavour violating processes. This is an interesting
phenomenology that one can study in this model.

It is described in section 2 that in the model of ref.\cite{ma-weg}, three
doublet and six triplet Higgses are proposed. The general form of the
scalar potential in this model can be written as
\begin{equation}
V=V_1(\Phi_i)+V_2(\xi_k)+V_3(\Phi_i,\xi_k)
\label{eq:pot}
\end{equation}
Here, $i=1,2,3$ and $k=1,\cdots,6$. The full terms in $V_1(\Phi_i)$, which
depend only on the three doublet Higgses, is given in ref.\cite{ma-raj}.
Terms in the scalar potentials of $V_2(\xi_k)$ and $V_3(\Phi_i,\xi_k)$
can be found in the following way. The general form of invariant scalar
potential under electroweak symmetry, containing
one doublet and triplet Higgses, is given in refs.\cite{pot-dt}. Now, this
potential needs to be genralised with three doublet and six triplet Higgses,
along with the imposition of the additional symmetry $A_4$. The resultant form
of that potential give full terms in $V_2(\xi_k)+V_3(\Phi_i,\xi_k)$. We
can notice here that in the full scalar potential of Eq.(\ref{eq:pot}),
there can exist may terms as compared to that in a model with one
doublet and triplet Higsses. Hence, we can expect lot more parameters
to be there in the scalar potential of ref.\cite{ma-weg}.
After minimizing the potential of Eq. (\ref{eq:pot}), $\Phi_i$ and $\xi_k$
get vevs, which need to satisfy certain relations in order to get consistent
neutrino mixing pattern in the model of ref.\cite{ma-weg}. The
minimization conditions for the part of
$V_1(\Phi_i)$ are studied in ref.\cite{ma-raj}. Now, the minimization
conditions for the scalar potential of Eq. (\ref{eq:pot}) can be studied,
and we believe, due to large number of parameters in $V$, these conditions
can be satisfied. One needs to know if this minima corresponds to local or
global minimum. We study these detailed topics in our future work.

In Eq. (\ref{eq:pot}), from the scalar potential part of
$V_3(\Phi_i,\xi_k)$, we can see that there are interaction terms between
doublet and triplet Higgses. This part of the potential can give mixing
masses between these two kind of Higgses, after $\Phi_i$ and $\xi_k$
acquire vevs. The vevs of these fields spontaneously break the electroweak
and $A_4$ symmetries of the model. After this breaking,
the following
fields remain in the theory: 6 doubly charged, 8 singly charged
and 17 neutral scalars. Out of these 17, 9 will be scalars and the rest
are pseudoscalars. One among the 9 neutral scalars
can be identified as the Higgs boson of SM. The masses of non-SM
scalars can be chosen to be around 1 TeV by appropriately choosing the
parameters of the scalar potential of Eq. (\ref{eq:pot}). Collider
signals of these scalars are briefly disscussed below. But before that,
from the interaction terms in the scalar potential of Eq. (\ref{eq:pot}),
we can notice that there can be tri-linear couplings involving one
neutral and two
charged scalars. These couplings may give additional contribution to Higgs
diphoton decay rate in the model of ref.\cite{ma-weg}. Since the measured
value related to this decay rate in the LHC experiment \cite{lhc} is
consistent with the SM prediction,
we may get some constraints on the above mentioned couplings of this model.

The doublet and triplet Higgses of this model have gauge interactions.
Moreover, they have Yukawa interactions with lepton fields.
Through these interaction terms, all the non-SM scalars of this model
can be produced at the LHC experiment through vector boson fusion and
subsequently they decay in to SM fields.
One can see that
the doubly charged scalars of this model can decay in to $\ell^\pm\ell^\pm$
and $W^\pm W^\pm$. Singly charged scalars of this model can decay in to
$\ell^\pm\nu$, $W^\pm Z$ and $W^\pm\gamma$. Neutral scalars of this model
can decay in to $\ell^+\ell^-$, $\nu\nu$, $W^+W^-$ and $ZZ$. If kinematically
allowed, through the interaction terms of Eq. (\ref{eq:pot}), a
doubly charged scalar can decay in to a pair of singly charged
scalars. We can notice here that an analysis of the collider signals
of this model is really interesting and worth to do.

\section{Conclusions}

In this work we have analysed a model which is proposed in ref.\cite{ma-weg}.
In this model neutrinos acquire masses and mixing pattern mainly due to the
presence of six Higgs triplets and $A_4$ symmetry. In order to explain
the mixing pattern among neutrinos, we have followed a certain approximation
procedure for diagonalising the neutrino mass matrix of this model.
We then have show that both NH and IH cases are possible for
neutrino masses in this model.
Following our approximation procedure, we have computed leading order
expressions for neutrino mixing angles and $\delta_{\rm CP}$. Using these
expressions we have shown that the current oscillation data can be explained
in this model.

\end{document}